\documentclass[reprint,superscriptaddress,aps,prl]{revtex4-2}

\usepackage{amsmath,amssymb,amsthm,mathrsfs}
\usepackage{graphicx}
\usepackage{dcolumn}
\usepackage{bm}
\usepackage{xcolor}
\usepackage{comment}
\usepackage{booktabs}
\usepackage{pifont}
\usepackage{tabularx}

\usepackage{lineno}

\usepackage[breaklinks,colorlinks,urlcolor=blue,linkcolor=blue,anchorcolor=blue,citecolor=blue]{hyperref}

\def\ii{{\rm i}}

\newcommand{\ex}[1]{\langle #1\rangle}

\def\hge{\hat{\sigma}_-}  
\def\heg{\hat{\sigma}_+} 
\def\hz{\hat{\sigma}_{z}}

\def\ga{\Gamma_{\rm 1D}}
\def\gp{\Gamma'}

\def\ra{\hat{\rho}}

\def\bra#1{\mathinner{\langle{#1}|}}
\def\ket#1{\mathinner{|{#1}\rangle}}

\def\opn{\hat{\mathcal{O}}_\nu}

\def\g2{g^{(2)}}

\def\sp{s^\perp}
\newcommand{\Jz}[1]{\langle\hat{J}_z^#1\rangle}
\newcommand{\Jcs}[1]{\langle\hat{J}_+^#1\hat{J}_-^#1\rangle}
\newcommand{\Jcd}[2]{\langle\hat{J}_+^#1\hat{J}_-^#2\rangle}

\newcommand\varpm{\mathbin{\vcenter{\hbox{%
  \oalign{\hfil$\scriptstyle+$\hfil\cr
          \noalign{\kern-.5ex}
          $\scriptscriptstyle({-})$\cr}%
}}}}

\begin{document}

\setlength{\parindent}{14pt}     
\setlength{\parskip}{0.5em} 

\title{Steady-state spin order and superradiance beyond the Dicke limit}

\author{Silvia Cardenas-Lopez}
 \affiliation{Department of Physics, Columbia University, New York, New York 10027, USA}
\author{Edgar Guardiola-Navarrete}
\affiliation{Department of Physics, Columbia University, New York, New York 10027, USA}
\author{Ana Asenjo-Garcia}
\email{ana.asenjo@columbia.edu}
\affiliation{Department of Physics, Columbia University, New York, New York 10027, USA}

\date{\today}

\begin{abstract}
Spontaneous collective decay in incoherently driven atomic ensembles can generate macroscopic coherence in the steady state, as exemplified by steady-state superradiance in single-mode cavities. Whether spontaneous order persists beyond the Dicke limit, where  competing collective decay channels and light propagation might preclude an ordered phase, remains an open question. We address it by analyzing incoherently pumped atoms coupled to one-dimensional electromagnetic baths through two models: a ring cavity with two bright decay channels, and a bidirectional waveguide where propagation additionally induces Hamiltonian dipole–dipole interactions. We find that both systems sustain steady-state phase order with intensity scaling as $N^2$,  but that the order takes two qualitatively distinct forms, neither described by a single macroscopic dipole. In the ring cavity, individual trajectories spontaneously break mirror symmetry, locking the atomic phases and the emitted field to one of two chiral orders. In the waveguide, coherent interactions instead enforce a phase-separated steady state in which the two chiral orders coexist, each dominating one end of the array. Our results show how competition and propagation shape emergent order beyond the Dicke limit.
\end{abstract}

\maketitle

The spontaneous emergence of order in systems composed of many interacting subsystems is pervasive across biology, chemistry, and the social sciences~\cite{Haken83}. In physics, order–disorder transitions have been extensively studied in thermodynamic equilibrium, yet some of the most striking manifestations occur far from equilibrium, where a featureless external energy input can stabilize robust ordered steady states and generate macroscopic coherence. A prime example is the conventional laser~\cite{Haken83,Haken84,Mandel95}, where atoms in a high-Q single-mode cavity, pumped above the lasing threshold, generate coherent light via stimulated emission. The intracavity field both quantifies the degree of organization and provides the feedback that enforces it.

Another example of spontaneous order in the form of macroscopic quantum coherence is that of a superradiant laser, realized in the bad-cavity limit where photons leak out faster than the time needed for stimulated emission~\cite{Meiser09,Meiser10,Zhang18,Liu20,Reilly25}. Here, the cavity acts as a shared electromagnetic reservoir that mediates all-to-all interactions between the atoms and, when the atoms are placed at the cavity antinodes,  endows the system with permutational symmetry. As a result, the atoms spontaneously decay via a single bright jump operator~\cite{Dicke54}. For suitable pumping strengths, the atoms reach a superradiant state~\cite{Dicke54}, emitting a field whose intensity scales quadratically with atom number and whose linewidth can fall below that of a single atom. The repeated action of the single jump operator synchronizes the atomic phases~\cite{Acebron05,Xu14,Zhu15}, establishing a macroscopic dipole resilient against quantum fluctuations. In contrast to a conventional laser, phase information is thus stored primarily in the atomic ensemble rather than in the cavity field.

An open question is whether steady-state spontaneous spin order survives when atoms decay into a multimode electromagnetic bath with continuous spectrum, beyond the single-mode Dicke limit.  In such settings, the competition between the resulting collective jump operators -- together with light-propagation effects encoded in the Hamiltonian -- may preclude the emergence of an ordered phase, or stabilize an order that cannot be described by a single macroscopic dipole. Prior work has shown that synchronization, characterized through a global order parameter, can survive in incoherently pumped dipolar arrays even in the presence of elastic interactions~\cite{Zhu15}. Separately, recent studies of \textit{transient} self-organization in extended systems -- including superradiant bursts~\cite{Masson20,Masson22,Sierra22,robicheaux2021theoretical,Liedl24,Rubies23}, directional photon emission~\cite{Cardenas23,Lukin25}, and spin ordering during decay from inverted states~\cite{Zhang25B} -- have revealed rich collective behavior during relaxation. Yet how competition and propagation shape the emergent order is not well understood. Two questions in particular remain open: whether the steady-state order in extended multimode reservoirs is captured by a single macroscopic dipole or develops nontrivial spatial structure, and whether it supports emission with superlinear scaling. 

Here, we address these questions in one-dimensional (1D) reservoirs, which provide a setting to isolate the roles of multimode competition and light propagation. We consider two spin models that progressively incorporate these effects: (i) a ring cavity~\cite{Lee25}, which supports two bright jump operators, each driving competing orders for the atomic coherence and emitted light, and (ii) a bidirectional waveguide, where in addition to competition between collective decay channels, light propagation generates coherent dipole–dipole interactions mediated by the waveguide. Beyond serving as minimal models to explore competition and propagation, these scenarios are directly realized in experimental platforms such as cold atoms coupled to microring resonators~\cite{Zhou24,Suresh25,Zhou25}, ring and bow-tie cavities~\cite{Cline25, Simon20, Chen22} and nanofibers~\cite{Vetsch10,Goban12,Gouraud15,Solano17}.

In the ring cavity and the waveguide, the emergent order takes two qualitatively different forms. In the ring cavity, two competing macroscopic orders maximize either left- or right-propagating photon emission, and in any individual realization a single order dominates, spontaneously breaking mirror symmetry. In the waveguide, by contrast, the two orders coexist in a phase-separated configuration, with chiral domains at opposite ends of the array separated by a domain wall. In both cases, the peak emission rate scales as $N^2$, matching the superlinear scaling of single-mode superradiant lasers. The upper threshold for collective emission is lower than in the Dicke limit, but a finite window of pumping strengths supports the ordered phase in both models.

\textbf{Theory--} We consider a system of $N$ two-level atoms, each with a resonance frequency $\omega_0$, interacting with a common electromagnetic reservoir at an individual decay rate $\ga$. Atoms also undergo parasitic individual decay at rate $\gp$ and are incoherently pumped at a rate $w$. After tracing out the electromagnetic field, the atomic dynamics under the Born-Markov approximation is governed by the master equation (making $\hbar=1$)~\cite{Gruner96,Dung02,Asenjo17}:
\begin{equation}
\dot{\ra}= -\ii\left[\hat{H},\ra\right]+\mathcal{L}_\text{dis}[\ra]+\mathcal{L}_\Gamma[\ra]+\mathcal{L}_w[\ra], \label{eq:master}    
\end{equation}
where 
\begin{subequations}
\begin{align}
\hat{H}&=\sum_{n,m=1}^N J_{nm}\heg^n\hge^m ,\label{Ham}\\
\mathcal{L}_\text{dis}[\ra]&=\sum_{n,m=1}^N\frac{\Gamma_{nm}}{2}\left(2\hge^m\ra\heg^n-\ra\heg^n\hge^m-\heg^n\hge^m\ra\right), \label{disipation}\\
\mathcal{L}_\Gamma[\ra]&=\sum_{n=1}^N\frac{\Gamma'}{2}\left(2\hge^n\ra\heg^n-\ra\heg^n\hge^n-\heg^n\hge^n\ra\right),
\end{align}
\end{subequations}
and the pumping term in the Lindbladian, $\mathcal{L}_w[\ra]$, is obtained from $\mathcal{L}_\Gamma[\ra]$ by swapping $\heg^n \leftrightarrow \hge^n $. Here, $\heg^n\equiv \ket{e_n}\bra{g_n}$ denotes the coherence operator for emitter $n$. The coefficients $\{J_{nm},\Gamma_{nm}\}$ encode the coherent and dissipative interactions, respectively, between atoms $n$ and $m$. These interactions are mediated by the shared electromagnetic environment and are fully determined by the Green's function of that environment, evaluated at the resonance frequency~\cite{Asenjo17}.

\begin{figure*}
  \centering
  \includegraphics[width=1\textwidth]{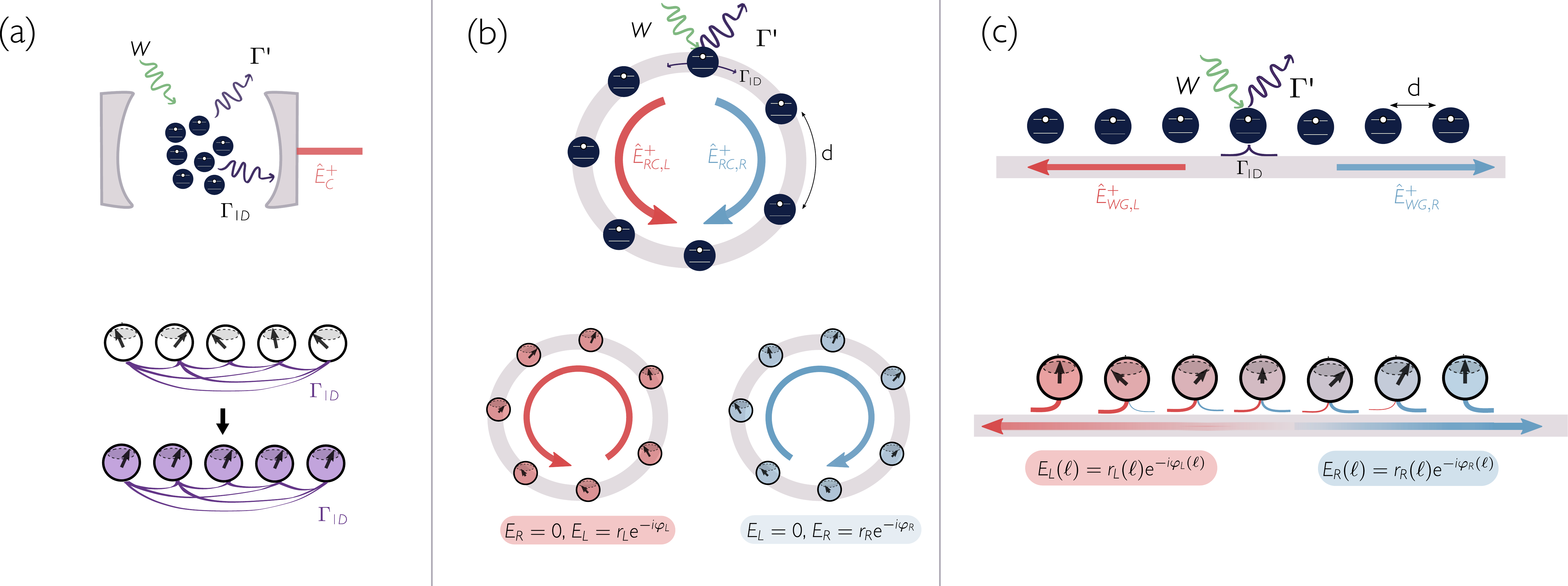}
\caption{\textbf{Schematics of the spin models -- single-mode cavity, ring cavity and waveguide -- and the spin orders they support}. In all settings, $N$ two-level atoms with parasitic decay $\Gamma'$ and incoherent pumping $w$ are coupled to a shared electromagnetic environment, into which they emit photons at rate $\Gamma_\text{1D}$. In the single-mode cavity (a), atoms provide global feedback to one another, and individual realizations of the steady state lock to a common phase. In the ring cavity (b), atoms provide competing directional feedback, and individual realizations lock to one of two macroscopic orders that emit all photons into either the right- or the left-propagating mode. In the bidirectional waveguide (c), feedback is directional and position-dependent, producing a steady state in which the two orders coexist on opposite sides of the array.}
\label{fig:Orders}
\end{figure*}

Collective dissipation, described by $\mathcal{L}_\text{dis}[\ra]$ in the master equation, can be recast in terms of atomic collective jump operators by diagonalizing the matrix of coefficients $(\boldsymbol{\Gamma})_{nm}\equiv \Gamma_{nm}$ as
\begin{equation}
\mathcal{L}_\text{dis}[\ra]=\sum_{\nu}\frac{\Gamma_\nu}{2}\left(2\opn\ra\,\opn^\dagger-\ra\,\opn^\dagger\opn-\opn^\dagger\opn\ra\right),
\label{eq:guided}
\end{equation}
where $\hat{\mathcal{O}}_\nu =\sum_{n=1}^N b_\nu^n \hge^n$ are collective jump operators with rates $\Gamma_\nu$ (and with spatial profile given by the eigenvectors $\bm b_\nu$ of $\boldsymbol{\Gamma}$). These modes may be superradiant ($\Gamma_\nu>\ga$, decaying faster than a single atom) or subradiant ($\Gamma_\nu<\ga$, decaying more slowly). Alternatively, $\mathcal{L}_\text{dis}[\ra]$ can be rewritten in terms of directional jump operators~\cite{Carmichael2000,Clemens2003} corresponding to left- and right-propagating emission, which we use throughout this work.

The steady state of the density matrix dynamics described by Eq.~\eqref{eq:master} is unique. This follows from Evans’ theorem~\cite{Evans77,Buca12}, which states that the steady state is unique whenever the dynamical generators of the master equation -- the Hamiltonian and all jump operators -- span the full operator space. In our case, this condition is met due to the presence of individual parasitic decay and incoherent pumping processes. Furthermore, the choice of incoherent pumping endows  Eq.~\eqref{eq:master} with global $U(1)$ symmetry, that is, invariance under the transformation $\heg^n\rightarrow e^{i \theta}\heg^n$ $\forall$ $n$. As a consequence of these two observations, the unique steady state must itself be $U(1)$-invariant, which implies that the expectation values of all emitter coherences, and hence all phase information, vanish identically in the steady state for any choice of initial state, in contrast to the case of coherent pumping. As we show throughout this work, this information can nonetheless be recovered by examining individual quantum trajectories or correlation functions of spin operators.

Signatures of order can also be observed at the density-matrix level in the intensity of the emitted electromagnetic field, which can be obtained from the atomic evolution via the input-output relations~\cite{Lalumiere13,Caneva15,Xu15_2}. The rate of photon emission into the common reservoir, proportional to the field intensity, reads
\begin{equation}
    R = \sum_{\nu}\Gamma_\nu \ex{\hat{\mathcal O}^\dagger_\nu\hat{\mathcal O}_\nu}.
\end{equation}
 Such signatures include transient superradiant bursts during decay from an inverted state~\cite{Dicke54}, and suppressed emission when subradiant states are populated~\cite{Albrecht19}. 

\textbf{Spin models -- }Figure~\ref{fig:Orders} illustrates the three spin models we consider: (a) atoms resonantly coupled to a bad cavity, whose photon leakage rate far exceeds all other timescales~\cite{Bonifacio71,Meiser10}; (b) atoms resonantly coupled to a ring cavity supporting a mode with wavenumber $k$; and (c) atoms coupled to a bidirectional waveguide supporting a guided mode with wavenumber $k$. The atomic dynamics in all three cases is governed by Eq.~\eqref{eq:master}, but the electromagnetic environment  differs across setups, giving rise to distinct atom-atom interactions and, consequently, different steady states. The interaction coefficients for the three models are respectively~\cite{Bonifacio71,Meiser10,Hummer13,Lee25,Asenjo17,Pichler15}:
\begin{align}
\text{Cavity:} \quad & J_{nm} = 0, \quad \Gamma_{nm} = \ga, \label{CavityCoefs} \\
\text{Ring cavity:} \quad & J_{nm} = 0, \quad \Gamma_{nm} = \ga \cos(k|z_n-z_m|), \label{eq:ringCavCoeffs} \\
\text{Waveguide:} \quad & J_{nm} - \ii \frac{\Gamma_{nm}}{2} = -\ii\frac{\ga}{2}\,e^{\ii k|z_n-z_m|}, \label{IntWg}
\end{align}
where $z_n$ denotes the position of atom $n$ along the ring cavity or waveguide, and $k$ is the wavenumber of the propagating light.

Atoms in a bad cavity realize the Dicke limit~\cite{Dicke54} with a single bright collective jump operator with collective decay rate $N\ga$. This is the only setting in which steady-state superradiance has been demonstrated~\cite{Bohnet12,Norcia16,Kristensen23} and provides our baseline for comparison. By contrast, both the ring cavity and the waveguide support two bright collective jump operators corresponding to left- and right- propagating emission~\cite{Pichler15},
\begin{equation}
    \hat{\mathcal{O}}_{L/R}=\frac{1}{\sqrt{N}}\sum_{n=1}^N e ^{\pm \ii k z_n}\hge^n,
\label{eq:left_right}
\end{equation}
each with a decay rate of $\Gamma_{L/R}=N\ga/2$.

In addition to these two competing jump operators, atoms coupled to a waveguide also have nonzero coherent interactions. For both the ring cavity and the waveguide, we assume that atoms form ordered arrays with spacing $d$, i.e., $z_n= nd$, though the results are expected to extend to disordered arrays since no argument is dependent on the spatial configuration. In the ``mirror'' configuration ($kd=n\pi$ with $n \in \mathbb Z$), the coherent part $J_{nm}$ vanishes, and the system supports a single bright mode, reducing the dynamics to that of a single-mode cavity~\cite{Chang12}. Expressions for the fields emitted in each geometry are given in the Methods section.

\textbf{Synchronization in the mean-field picture --} A mean-field treatment provides analytical insight into the emergent spin order by casting it as a phase-synchronization problem. Within this approximation, each atom is treated as a classical dipole described by an inversion, a coherence amplitude, and a phase, $\{s_n^z, s_n^\perp, \phi_n\}$, defined by
\begin{subequations}
\begin{align}
    2s^z_n &\equiv \langle\hat{\sigma}_z^n\rangle, \\
    \sp_n e^{i\phi_n} &\equiv \langle\heg^n\rangle,
\end{align}
\end{subequations}
with the corresponding equations of motion derived from a factorized density matrix $\rho(t)\approx\bigotimes_{n=1}^N\rho_n(t)$ (see Methods). The phase dynamics takes the form
\begin{equation}
\dot{\phi}_\ell=\frac{s^z_\ell}{\sp_\ell}\sum_{n\neq \ell}\sp_n \left(-2 J_{\ell n}\cos(\phi_{n \ell})+\Gamma_{\ell n}\sin(\phi_{n \ell})\right),
\label{MeanFieldphasesMainText}
\end{equation}
where $\phi_{n\ell}\equiv \phi_n-\phi_\ell$. This expression maps onto the Sakaguchi--Kuramoto model~\cite{Zhu15,Sakaguchi86}, a paradigmatic description of synchronization and pattern formation in networks of classical limit-cycle oscillators~\cite{Acebron05}. The two coupling terms have distinct origins and effects: the cosine term comes from the coherent dipole--dipole interactions $J_{\ell n}$, while the sine term comes from the dissipative collective decay $\Gamma_{\ell n}$. 

  \begin{figure*} \centering \includegraphics[width=1\textwidth]{Figures/MeanFieldRingCavityAlt_compressed.pdf} \caption{\textbf{Order in the ring cavity.} Mean-field trajectories  and Wigner function in (a) the space of right- and left-field amplitudes, and (b) in the complex plane of the right field. Diamonds indicate  final configurations of selected mean-field trajectories. Time is encoded in the line transparency (fainter for earlier times), and each trajectory is colored according to which field dominates in the steady state.  In (b), the Wigner function in regions far from the origin has been multiplied by a factor of 10 for visualization purposes. Panel (c) shows the magnetization $\mathcal{M}(n, t)$ for a single mean-field realization in which the right field prevails. Parameters: $N=1024$, $\Gamma' = 0$, $w = N\ga/4$.} \label{fig:RingCavMF} \end{figure*}
  
Evaluating the input–output relation~\cite{Asenjo17,Caneva15} at the mean-field level, we can rewrite the phase dynamics as
\begin{equation}
\dot{\phi}_\ell=\frac{2s^z_\ell}{\sp_\ell}\text{Re}\left[e^{\ii \phi_\ell }E(z_\ell)\right],
\label{MeanField_field}
\end{equation}
where $E(z_\ell)$ is the field emitted by the whole array at the position of atom $\ell$. The contribution of atom $\ell$ itself is in quadrature with its dipole and exerts no torque on its phase, so the synchronizing agent is the light emitted by the rest of the ensemble: each atomic phase is pulled toward alignment with the local field, as
in injection locking. While the underlying physical picture dates back to semiclassical treatments of transient superradiance~\cite{Gross82}, Eq.~\eqref{MeanField_field} distills it into a single phase equation,
applicable to any electromagnetic environment and, in our case, to the pumped steady state. As we show below, mode competition and light propagation determine whether and how the atomic phases synchronize in our three spin models.
  
For atoms in a bad cavity, the mean-field phase equation takes the form already discussed in Refs.~\cite{Meiser09,Meiser10,Xu14,Zhu15},
\begin{equation}
\dot{\phi}_\ell=\frac{\ga s_\ell^z}{\sp_\ell}r \sin(\psi-\phi_\ell), \label{AngularEqCavityMeanField}
\end{equation}
where \(r\) and \(\psi\) are the amplitude and phase of the field emitted by the cavity, $E=\ii \frac{\ga}{2} r e^{-\ii \psi}$. The cavity field expectation value thus plays the role of the Kuramoto order parameter. Within the synchronization window, this realizes a Kuramoto model with positive coupling, and the atoms relax to a steady state with well-defined field with amplitude $r$ and an arbitrary global phase $\psi$, reflecting the global $U(1)$ symmetry of the master equation. Quantum fluctuations induce phase diffusion of the mean-field solution~\cite{Mandel95,Tieri17}, and  shot-to-shot the phase is picked by spontaneous symmetry breaking, as in conventional single-mode lasers. 
  
  \textbf{Ring cavity: winner takes all --} In contrast to the single-mode case, we find that the ring cavity supports two competing collective decay channels, and their competition spontaneously breaks both the global $U(1)$ symmetry and the mirror symmetry between left- and right-propagating orders at the level of individual realizations. Since the field at any position decomposes into left- and right-propagating components, we define classical order parameters for left and right emission with amplitudes $r_{L/R}$ and phases $\psi_{L/R}$, proportional to the respective fields, as
 \begin{equation} E_{L/R}= \sum_{n=1}^N e^{\pm i k z_n}\sp_n e^{-\ii \phi_n} \equiv r_{L/R} e^{-\ii \psi_{L/R}}.\label{OrderParamsRing} \end{equation} 
The synchronization dynamics~\eqref{MeanField_field} then becomes
  \begin{equation} \dot{\phi}_\ell=\frac{\ga s_\ell^z}{2\sp_\ell}\sum_{\nu= \text{L,R}} r_\nu \sin(\psi_\nu-\theta_{\nu,\ell}-\phi_\ell), \label{AngularEqRingMeanField} \end{equation} 
  where we have defined the position-dependent phases $\theta_{L/R,\ell}=\mp k z_\ell$. When pumping produces positive inversion and nonzero coherences, the dynamics involves two competing feedback loops that pull $\phi_\ell$ toward either the right or the left order, i.e.,
  \begin{equation}
      \phi_\ell^{R} = \psi_{R} - k z_\ell, \qquad \phi_\ell^{L} = \psi_{L} + k z_\ell.
      \label{orders}
  \end{equation}Each attractor maximizes either $r_R$ or $r_L$, and the strength of attraction grows with the corresponding field amplitude. This creates a positive feedback loop: whichever field is momentarily larger attracts phases toward its fixed point, amplifying itself further and suppressing the rival mode. As captured by Eq.~\eqref{OrderParamsRing} and sketched in Fig.~\ref{fig:Orders}(b), the light emitted by each atom into the right (left) mode serves as feedback to synchronize the rest of the atoms to the right (left) order.

  Within the synchronization window, as shown in Fig.~\ref{fig:RingCavMF}(a) and (b),  mean-field trajectories relax to one of two steady states: either the left field acquires a finite amplitude while the right field is suppressed, or vice versa, reminiscent of conventional ring-laser behavior~\cite{Mandel95}. To visualize the microscopic phase order, we define the ``magnetization'' for atom $n$ at time $t$ as
  \begin{equation} \mathcal{M}(n,t)=\frac{|\Delta\phi_n(t)-kd|-|\Delta\phi_n(t)+kd|}{2 kd}, \label{magnetization}\end{equation}
  where $\Delta\phi_n\equiv \phi_n -\phi_{n-1}$. The magnetization  quantifies whether neighbors are closer to maximizing right ($\mathcal{M}(n,t)= 1$) or left ($\mathcal{M}(n,t)= -1$) emission. Figure~\ref{fig:RingCavMF}(c) shows that, after an initial transient, all the phases stabilize to either the right or the left order. This mean-field picture is qualitatively confirmed by simulations including quantum fluctuations, which we discuss next.

  Beyond mean field, we use the  Truncated Wigner Approximation (TWA)~\cite{Mink22,Mink23,Guardiola25} to include leading quantum fluctuations (see Methods). The fluctuations broaden the mean-field solutions into a bimodal Wigner distribution concentrated near the two attractors, as shown in Fig.~\ref{fig:RingCavMF}(a). The Wigner function of the right field [Fig.~\ref{fig:RingCavMF}(b)] features both a concentration at the origin and an outer ring at finite amplitude. The weight at the origin corresponds to realizations in which the left order dominates and the right field is suppressed; the outer ring corresponds to realizations in which the right order dominates, with phase undetermined on average due to the $U(1)$ symmetry. The left field displays the mirror-image structure.

The steady state of Eq.~\eqref{eq:master} is unique, as established in the Theory section. The two ordered configurations therefore do not correspond to distinct steady states of the master equation, but to two dynamical outcomes that the system switches between at the level of individual trajectories. The mirror-inversion symmetry of Eq.~\eqref{eq:master} for the ring cavity is a weak symmetry~\cite{Buca12,Albert14} --  preserved at the level of the master equation but not at the level of single trajectories. Consistent with this, the bimodal structure of  the Wigner distribution results from stochastic switching between the two dynamical outcomes in measurement-conditioned trajectories~\cite{Wiseman09}, in contrast to dissipative freezing that occurs in the presence of a strong symmetry~\cite{Sanchez19}. As the atom number increases, the two lobes of the Wigner distribution become increasingly sharp and well-separated, so that any individual trajectory is localized near one of the two attractors. Each experimental realization therefore effectively breaks mirror symmetry, even though the ensemble-averaged steady state remains symmetric.
    
 The locking to either a global left or right phase order motivates an ansatz for the steady state as an incoherent superposition of spin-coherent states with phases matching the two possible orders~\cite{Zhang25,Zhang25B}:
\begin{equation}
    \hat{\rho}_\text{ans,RC}(t_{ss}) = \frac{1}{4\pi}\int_0^{2\pi}d\psi\left[\hat{\rho}_R(\psi)+\hat{\rho}_L(\psi)\right],
    \label{eq:ansatzRingCavity}
\end{equation}
where $t_{ss}$ denotes any time at which the system has reached its steady state, and
\begin{eqnarray}
    \begin{aligned}
        \hat{\rho}_{L/R}(\psi) &=\bigotimes_{n=1}^N\ket{\theta_n,\pm kz_
        n+\psi}\bra{\theta_n,\pm kz_n+\psi}.
    \end{aligned}
\end{eqnarray}
The spin coherent states are defined as $ \ket{\theta,\phi} = \cos \frac{\theta}{2}\ket{e}+ e^{\ii \phi}\sin \frac{\theta}{2}\ket{g}$, and the free angle $\psi$ in the ansatz reflects the global $U(1)$ symmetry. Although the steady state of Eq.~\eqref{eq:master} has vanishing atomic coherences due to the $U(1)$ symmetry, measurement-conditioned trajectories develop nonzero coherences via measurement backaction. Heterodyne unraveling~\cite{Wiseman09} is particularly well-suited for this purpose: each realization develops a definite phase and, after sufficiently long evolution, tends to synchronize to either the left or the right order.

Observables sensitive to the synchronization direction vanish on average, since roughly half of the realizations lock to each side. Nonetheless, signatures of the order are observable at the level of correlation functions~\cite{Zhu15}
\begin{equation}
C_{nm}=\langle\heg^n\hge^m\rangle(t=t_{ss}).\label{correlations}
\end{equation}
Evaluating the above equation using $\hat{\rho}_\text{ans,RC}(t_{ss})$, the equal-weight superposition of left- and right-ordered configurations produces a cosine spatial pattern
\begin{equation}
\begin{aligned}
    C_{nm}& \propto \frac{1}{2}\left(e^{-\ii k(z_n-z_m)}+e^{\ii k(z_n-z_m)}\right)\\&= \cos(k(z_n-z_m)),
    \label{CorrRing}
\end{aligned}
\end{equation}
which is in agreement with the behavior observed in our TWA simulations.
    
    \begin{figure} \centering \includegraphics[width=1\columnwidth]{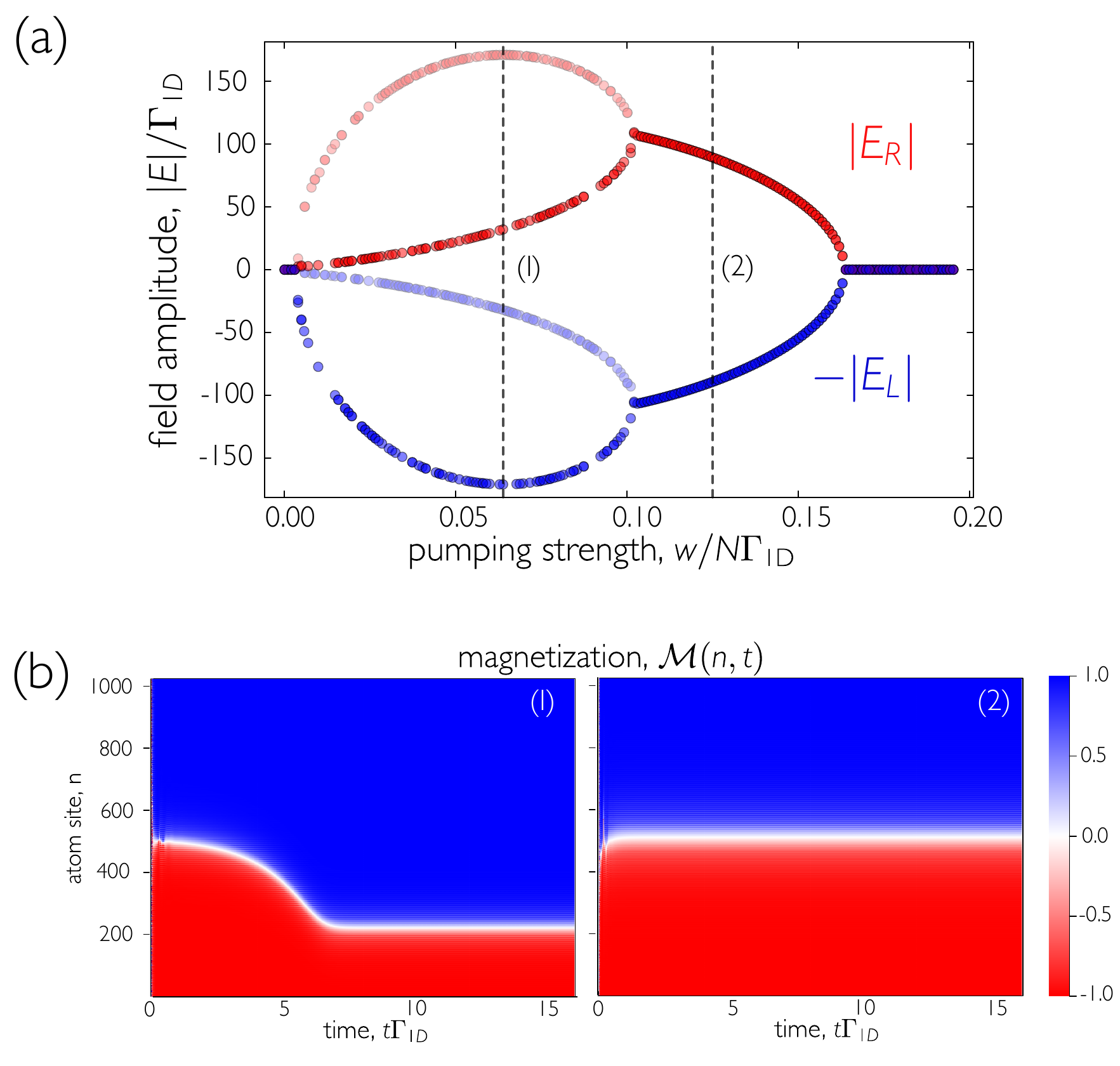} \caption{\textbf{Domain wall behavior in bidirectional waveguide.} (a) Mean-field amplitudes of the left and right order parameters. Below $w\simeq 0.1 N\ga$, two solutions coexist (distinguished by point transparency): one in which the right field dominates (opaque) and one in which the left field dominates (transparent). (b) Mean-field magnetization $\mathcal{M}$ for two realizations of the phase dynamics at the pumping rates indicated in (a): (left) the domain wall is displaced from the center ($w=0.064N\ga$) and (right) it stabilizes at the center ($w=N\ga/8$). In all plots, $(N,\Gamma',kd) = (1024,2\ga,2\pi/3)$.} \label{fig:LobePLot} \end{figure}

\textbf{Waveguide: coexistence of left and right orders --} Besides collective decay into two competing decay channels, the waveguide mediates coherent atomic interactions that encode light propagation~\cite{Gross82,Goncalves25}. Light propagation makes feedback position-dependent, as light emitted by atom $\ell$ to the right-propagating mode only serves as feedback for atoms located to its right, as sketched in Fig.~\ref{fig:Orders}(c). This position-dependence is reflected in the phase dynamics,
\begin{equation} \dot{\phi}_\ell=\frac{\ga s_\ell^z}{\sp_\ell}\sum_{\nu= \text{L,R}} r_\nu(\ell) \sin(\psi_\nu(\ell)-\theta_{\nu,\ell}-\phi_\ell), \label{AngularEqWQEDMeanField} \end{equation}
where the feedback synchronizing atom $\ell$ to the right or left is given by local order parameters proportional to the fields evaluated at $z_\ell$, i.e.,
   \begin{subequations}
   \label{OrderParamWg}
\begin{align}
 E_R(\ell)&=\sum_{z_n<z_\ell}e^{- \ii k z_n}\sp_n e^{-\ii \phi_n}\equiv r_R(\ell) e^{-\ii \psi_R(\ell)},\label{LocalOrderR}\\ E_L(\ell)&=\sum_{z_n>z_\ell}e^{\ii k z_n}\sp_n e^{-\ii \phi_n}\equiv r_L(\ell) e^{-\ii \psi_L(\ell)}.\label{LocalOrderL}\end{align}\end{subequations}

 Unlike the ring cavity [Fig.~\ref{fig:RingCavMF}(c)], where a single macroscopic order dominates each realization, atoms in a bidirectional waveguide phase-separate between the two competing orders. The location of the domain wall between the two ordered phases depends on the pumping rate. At low pumping rates, there is bistability [Fig.~\ref{fig:LobePLot}(a)], with either the right- or left-propagating field dominating and the domain wall displaced from the center, whereas at high pumping rates the domain wall is instead pinned at the center. Emission into the right (left) mode accumulates along the array, and peaks at the right (left) edge, so atoms near the edges are more coherent and more strongly locked into order. The impact of coherent interactions is made evident in the magnetization $\mathcal{M}(n,t)$, shown in Fig.~\ref{fig:LobePLot}(b) for two mean-field trajectories. 

\begin{figure} \centering \includegraphics[width=1\columnwidth]{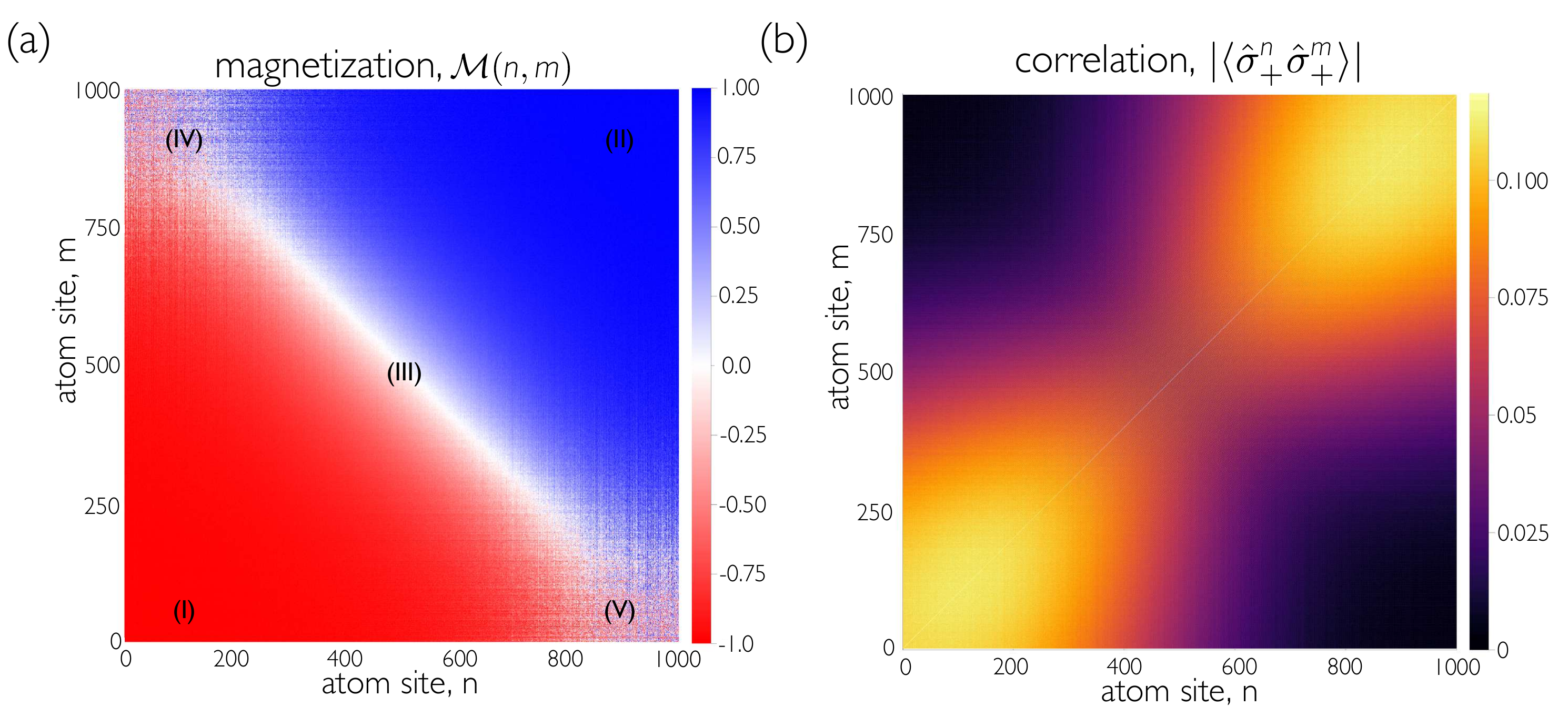} \caption{\textbf{Quantum fluctuations and spin order in a bidirectional waveguide.} (a) Pairwise magnetization $\mathcal{M}(n,m)$ and (b) absolute value of the correlation function $C_{nm}=\langle\heg^n\hge^m\rangle$ in the steady state, obtained from TWA simulations.  In (a), labels (I)–(V) denote regions of distinct behavior discussed in the main text. In (b), the color scale is saturated at $~0.1$ for visualization. In both plots, $(N,\Gamma',w,kd) = (1024,2\ga,N\ga/8,2\pi/3)$.} \label{fig:WgMF} \end{figure}

The steady-state coherence matrix $C_{nm}$, defined in Eq.~\eqref{correlations}, shows that this order persists in the presence of quantum fluctuations. We generalize the magnetization defined in Eq.~\eqref{magnetization} to a matrix comparing the correlation phases $\arg[C_{nm}]$ with the phase difference expected from left- and right- ordered states [Eq.~\eqref{orders}] by defining $\theta^\pm_{nm}(t)=\arg[C_{nm}(t)e^{\pm\ii(n-m)kd}]$ and

\begin{equation}
\mathcal{M}(n,m,t)
= \frac{|\theta^-_{nm}(t)|-|\theta^+_{nm}(t)|}{|\theta^-_{nm}(t)|+|\theta^+_{nm}(t)|},
\end{equation}
so that $\mathcal{M}(n,m,t)\simeq \pm 1$ when the phase difference is consistent with right- (left-) ordered states. The steady-state magnetization is shown in Fig.~\ref{fig:WgMF}(b). Atoms at the edges consistently lock to either the left or the right mode, resulting in $\mathcal{M}(n,m,t_{ss})\approx \pm 1$ for correlations between atoms within the same chiral domain [regions (I) and (III)]. 

A full understanding of the phases of the coherence matrix requires considering that 
$C_{nm}$ is an average over quantum trajectories, and fluctuations can shift the global phase of each domain or displace the domain wall between realizations. For instance, at large pumpings, atoms near the center of the array synchronize with the left  or right mode with roughly equal probability, so that their pairwise correlations [region (II)] average to $C_{nm}\simeq \cos(k(z_n-z_m))$, as in the ring cavity [Eq.~\eqref{CorrRing}]. Finally, pairs of atoms belonging to different domains [regions (IV) and (V)] have vanishing correlations, as shown in Fig.~\ref{fig:WgMF}(b), suggesting that the relative phase between the two chiral domains fluctuates freely across realizations, washing out interdomain correlations upon averaging.

In the spirit of Eq.~\eqref{eq:ansatzRingCavity}, we condense this information in an ansatz for the steady state as an incoherent superposition of spin-coherent components.
\begin{equation}
\begin{aligned}
    &\hat{\rho}_\text{ans,WG}(t_{ss}) = \int_0^{2\pi}\frac{d\psi_R}{2\pi}\int_0^{2\pi}\frac{d\psi_L}{2\pi}\sum_{\textbf{A}\in\{R,L\}^N}P(\textbf{A})\\&\times\bigotimes_{n=1}^N \ket{\theta_n,\phi(A_n,n,\psi_{A_n})}\bra{\theta_n,\phi(A_n,n,\psi_{A_n})},
    \end{aligned}
    \label{eq:ansatzWaveguide}
\end{equation}
with
\begin{equation}
    \phi(A_n,n,\psi_{A_n}) = \begin{cases}
        kz_n +\psi_L&\text{  if  } A_n=L\\
        -kz_n +\psi_R&\text{  if  } A_n=R.
    \end{cases}
\end{equation}
Here, $P(\mathbf{A})$ denotes the probability of a given configuration of local orderings. For example, $\mathbf{A} = [L,L,R,L\ldots]$ means that the first two atoms align with the left order, the third aligns with the right order, and so on. The independent integration over $\psi_R$ and $\psi_L$ reflects the global $U(1)$ symmetry and the assumption that the relative phase between the left and right ordered domains is completely undetermined. By proposing a probability model for the configurations $P(\mathbf{A})$ that captures the tendency of edge atoms to lock to the left and right order, we quantitatively reproduce the characteristic structure of the correlation matrix $C_{nm}$ with few fitting parameters. Details on the probability model, fitting parameters, and correlation calculation are provided in the SI.

\textbf{Intensity of the emitted light --} We determine the pumping thresholds for synchronization and superradiant emission, and show that the emitted intensity scales as $N^2$ for all three spin models. In a cavity, the lower and upper thresholds for superradiant lasing are known to be $w^{(l)}=\gp$ and $w^{(u)}=N\ga$, respectively, with a maximum emission rate scaling as $R_\text{max} = N^2\ga/8$~\cite{Meiser09,Meiser10}. The thresholds are intuitively clear since incoherent pumping must exceed the independent decay rate to generate positive inversion and enable synchronization, while excessively strong pumping destroys correlations seeded by collective decay. Since $N\ga$ sets the maximum collective decay timescale, for $w>w^{(u)}$ correlations are washed out faster than they can form. 

For the ring-cavity, we find analytical expressions for the thresholds for superradiant emission by exploiting the superspin method introduced in Ref.~\cite{Lee25}. Analyzing the Heisenberg evolution of the collective operators and performing a second-order cumulant expansion analogous to the single-mode case (see the SI), we find that the ring-cavity steady state also exhibits threshold behavior. In particular, the steady-state total intensity and excited–state population $P_e\equiv\sum_n\langle\hat{\sigma}_{ee}^n\rangle$ read
\begin{subequations}
\label{ThresholdsRing}
\begin{align}
       R(t=t_{ss})&=\begin{cases}
            0  \text{     if    } w\notin [\gp,N\ga/2] \\ 
            \begin{aligned}
                \frac{wN}{2}\left(1-\frac{\gp}{w}-\frac{2}{w}\frac{(w+\gp)^2}{N \ga}\right)\\[0.7em]
            \text{     if    } w\in [\gp,N\ga/2],
            \end{aligned}
        \end{cases}
        \label{sol1SL}\\
         P_e(t=t_{ss})&=\begin{cases}
            \frac{(w-\gp)N}{2(w+\gp)} +\frac{N}{2}  \text{     if    } w\notin [\gp,N\ga/2] \\ 
            \frac{(w+\gp)}{\ga}+\frac{N}{2} \text{     if    } w\in [\gp,N\ga/2].
        \end{cases}
        \label{sol2SL}
\end{align}
\end{subequations}

\begin{figure*} \centering \includegraphics[width=\textwidth]{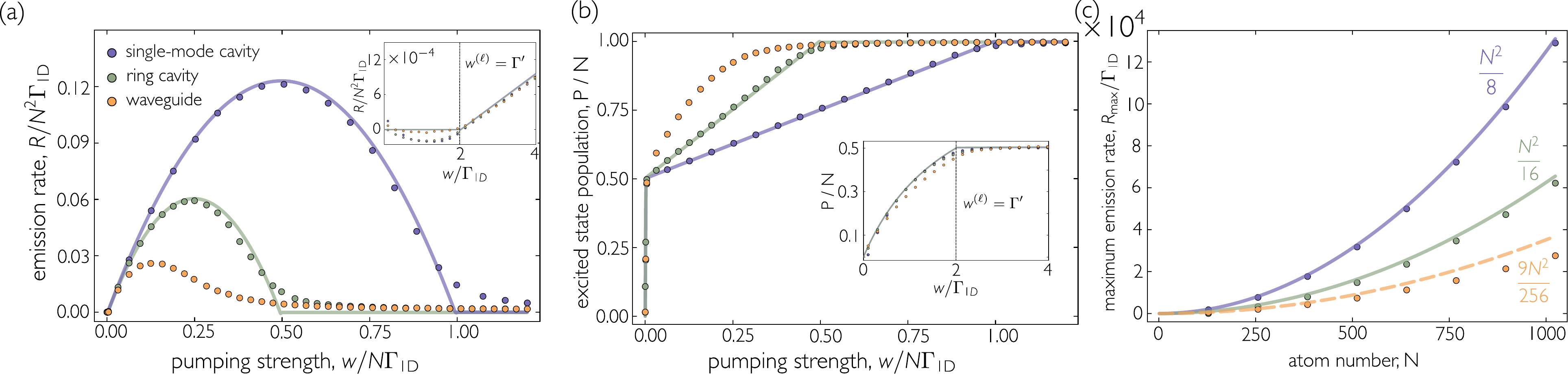} \caption{\textbf{Thresholds for superradiant light emission and scaling of maximum emission rate}. (a) Emission rate and (b) excited-state population in the steady state as a function of pumping strength. Insets: zoom on the first threshold. (c) Maximum intensity vs atom number. In all plots, solid lines show the analytical expressions [from Ref.~\cite{Meiser10} for the single-mode cavity and from Eqs.~\eqref{ThresholdsRing} for the ring cavity], points show the numerical results with TWA, and the dashed line in (c) indicates the estimate of $R_{\text{max}}$ for the waveguide described in the main text.  Parameters: $(N,\gp,kd )=(1024,2\ga,2\pi/3)$.} \label{fig:Thresholds} \end{figure*}

Figures~\ref{fig:Thresholds}(a,b) show good agreement between these expressions and the values obtained with TWA. Numerical results for atomic spacings outside the superspin regime follow the same behavior, indicating that the steady-state values of these observables are largely insensitive to the spatial arrangement of the atoms. Equations~\eqref{ThresholdsRing} predict sharp changes in population and the onset and quenching of steady-state collective emission at a lower threshold $w^{(\ell)}=\Gamma'$ and an upper threshold $w^{(u)}=N\Gamma_\text{1D}/2$. The latter is half the single-mode cavity value, reflecting the splitting of collective emission between the two directional channels. Since each jump operator $\hat{O}_{L/R}$ acts at a collective decay rate $N\ga/2$, correlations seeded by either channel are washed out at half the pump strength required in the single-mode case. The insets of Fig.~\ref{fig:Thresholds}(a,b) zoom into the behavior around the first threshold. Below $w^{(\ell)}$, TWA yields a small unphysical negative decay rate, consistent with the fact that the steady state in this regime is known to be dark and highly entangled for a single-mode cavity~\cite{Shankar21}, a scenario expected to carry over to 1D reservoirs. Despite this limitation, TWA correctly captures the location of $w^{(\ell)}$. Moreover, Eq.~\eqref{sol1SL} predicts a maximum emission rate at $w=N\Gamma_\text{1D}/4$, with $R_{\text{max}} = N^{2}\Gamma_\text{1D}/16$, in agreement with numerical results as shown in Fig.~\ref{fig:Thresholds}(c). As for a single-mode cavity, the upper threshold reflects the requirement that the pump strength remain below the maximum dissipation rate of the system~\cite{Meiser10,Mok24}, which also sets the fastest synchronization timescale in Eq.~\eqref{AngularEqRingMeanField}.

For the waveguide, we obtain the thresholds from TWA simulations, since the mean-field equations capture the steady-state intensities only qualitatively. We find $w^{(\ell)}\sim \Gamma'$ for all array spacings, but the exact position of the upper threshold depends on lattice constant. Away from the ``mirror'' configuration, we estimate the upper threshold from Eqs.~\eqref{OrderParamWg}. Because the feedback is position-dependent, each atom has a different maximum pumping at which it can remain synchronized. For instance, atom 1 experiences feedback from $(N/2-1)$ atoms driving it toward the left order, so the maximum pumping strength for synchronization is $w^{(u)}(n=1)\sim (N/2-1)\ga$. More generally, atom $n$ in the first half of the array will synchronize only if $w < w^{(u)}(n) \sim \left(N/2-n\right)\ga$. Averaging over the array then yields $\bar{w}^{(u)} \sim N\ga/4$, and by similar reasoning, an optimal pumping $\bar{w}^{(opt)} \sim N\ga/8$. Counting atoms with $w^{(u)}(n)>\bar{w}^{(opt)}$ that synchronize to either order at the optimal pump (denoted $N_\text{sync}$) and using the known single-mode cavity scaling, we estimate the waveguide intensity at optimal pumping to be $R_\text{max} \sim 2\left(N_\text{sync}^2\ga/8\right) =9N^2 \ga/256$. These estimates qualitatively agree with the numerical results in Fig.~\ref{fig:Thresholds}. In contrast to transient superradiance, where coherent interactions can often be neglected when analyzing the superradiant burst~\cite{Masson22,Cardenas23}, the coherent interactions intrinsic to the waveguide strongly modify the properties of the radiation emitted in the steady state.

A second notable property of the light emitted by steady-state superradiance in a single-mode cavity is its narrow spectrum, whose linewidth approaches $(\Delta\nu)_\text{min}\sim\ga$ in the large-$N$ limit~\cite{Meiser09}. Whether a comparably narrow linewidth can be achieved in extended systems remains an open question~\cite{Bychek25}; we address this point and present a preliminary numerical analysis in the SI.

\textbf{Conclusions and outlook -- }We have studied steady-state superradiance of incoherently pumped atoms coupled to 1D reservoirs, focusing on two setups: a ring cavity, where atoms collectively decay into left- and right-propagating modes, and a bidirectional waveguide, where coherent exchange processes accompany collective decay. In the ring cavity, individual realizations lock to one of two macroscopic orders that maximize emission into a single direction, whereas in the waveguide coherent interactions enforce coexistence of left and right order, with edge atoms synchronizing to opposite modes. This self-organization shapes the emitted light. For the ring cavity we derived analytical thresholds and optimal intensities; for the waveguide, we obtained numerical estimates of the corresponding quantities.

Our results suggest several directions for future work. On the experimental side, they motivate studies of steady-state superradiance in ring resonators, bow-tie cavities, and nanofibers. For the waveguide, experiments are especially needed to validate the theory, since our different approaches agree qualitatively but differ in their quantitative predictions. On the theoretical side, a more exhaustive characterization of the statistics and spectrum of the emitted light from these ordered states would be valuable. It would also be worthwhile to quantify shot-to-shot fluctuations in the ordering through semiclassical simulations of individual trajectories~\cite{Yu24,Zhang25}, which could clarify the stability of the phase-ordered state and shed light on the origin — or possible absence — of linewidth narrowing and coherence in the emitted field.

Finally, a natural extension of this work is to explore the prospects for superradiant lasing in free space, as originally proposed by Dicke~\cite{Dicke64}. In particular, it would be interesting to see whether collective decay in the absence of coherent interactions generically leads to basins of attraction associated with the brightest collective modes, as in the ring cavity, and whether the coherent interactions generate crossovers between competing directional orders, as observed in the waveguide. Recent work~\cite{Bychek25} has shown that a few atoms in a 1D array can emit light with a linewidth that converges to the natural single-atom value by limiting pump-induced broadening and driving only a subset of atoms. In practice, however, 1D atomic arrays with ultranarrow transitions may not be suitable to generate light for atomic clock interrogation, since their maximum collective decay rate scales only linearly with $N$~\cite{Mok24}, preventing the realization of superradiant lasers (as the emission is too weak even for large systems). Nevertheless, it would be worth investigating whether selectively pumping only part of a higher-dimensional array could overcome this limitation and enable superradiant lasing in free-space geometries.

\textit{Acknowledgments --} We are grateful to Joseph T. Lee, Ricardo Gutierrez-Jauregui, Xueyue Zhang and Tingyao Wang for discussions. We acknowledge support by the National Science Foundation through the CAREER Award (No. 2047380), the Air Force Office of Scientific Research through their Young Investigator Program (grant No. 21RT0751), the David and Lucile Packard Foundation, and the Chu Family Foundation.

\section{Methods}
\subsection*{Zero- and one-dimensional reservoirs}

Here, we provide additional details for the three spin models. 

For a single-mode cavity, the single bright collective jump operator is
\begin{equation}
    \hat{\mathcal{O}}_C=\frac{1}{\sqrt{N}}\sum_{n=1}^N \hge^n.
    \label{DickeOp}
\end{equation}
The positive frequency component of the field operator is directly related to the jump operator and reads~\cite{Bonifacio71} 
\begin{equation}
    \hat{E}_C^+ = \ii\frac{\ga}{2}\sum_{n=1}^N\hge^n.
    \label{fieldCavity}
\end{equation}

 For the ring cavity, using the input-output relation~\cite{Caneva15}, the left- and right- propagating fields at any position $z$ are
\begin{subequations}
\begin{align}
    \hat{E}_{RC,R}^+(z) &=\ii\frac{\ga}{4}\sum_{n=1}^N e^{\ii k(z-z_n)}\hge^n, \\
    \hat{E}_{RC,L}^+(z) &=\ii\frac{\ga}{4}\sum_{n=1}^N e^{\ii k(z_n-z)}\hge^n, 
    \end{align}
\end{subequations}
which are proportional to the operators $\hat{\mathcal{O}}_{L/R}$. 

Finally, for a bidirectional waveguide the input-output relation yield the following expressions for the right- and left- propagating emitted fields at any position $z$ along the waveguide 
\begin{subequations}
\begin{align}
    \hat{E}_{WG,R}^+(z) &=\ii\frac{\ga}{2}\sum_{z>z_n} e^{\ii k(z-z_n)}\hge^n, \\
    \hat{E}_{WG,L}^+(z) &=\ii\frac{\ga}{2}\sum_{z<z_n} e^{\ii k(z_n-z)}\hge^n. 
    \end{align}
\end{subequations}
When the observation point $z$ lies far to the right (left) of all the atoms, the expression for $\hat{E}_{WG,R}^+(z)$ ($\hat{E}_{WG,L}^+(z)$) reduce, up to a global phase, to the right (or left) jump operator of Eq.~\eqref{eq:left_right}.

\subsection*{Analytical and computational methods}
 When full permutation symmetry is present (as in a single-mode cavity) the dynamics can be computed exactly with computational cost $\sim N^3$~\cite{Xu13}. This can be extended to ring cavities by exploiting partial permutation symmetry in commensurate ``superspin'' arrays~\cite{Lee25}. In particular, for spacings $d= m\lambda_0 /(2p)$ with $m,p\in\mathbb{Z}$, the ensemble decomposes into $p$ symmetric groups $\mathcal{S}_\alpha$ ($\alpha=1,\dots,p$), each represented by a collective spin
\begin{subequations}
\label{eq:superspins}
\begin{align}
     \hat{J}^\alpha_-=&\frac{1}{N_\alpha}\sum_{n\in \mathcal{S}_\alpha} c_{n,\alpha}\hge^n,\\
    \hat{J}^\alpha_z=&\frac{1}{2N_\alpha}\sum_{n\in \mathcal{S}_\alpha} \hz^n,
\end{align}
\end{subequations}
where $N_\alpha$ is the number of atoms in $\mathcal{S}_\alpha$ and the coefficients $\{c_{n,\alpha}\}$ ensure that coherences of atoms within the same symmetric group add in phase (see SI for more details). This reduces the computational complexity to $\sim N^{3p}$ and enables a compact description fully in terms of the collective operators in Eqs.~\eqref{eq:superspins}. The case $p=1$ reduces to the single macroscopic dipole describing a superradiant laser~\cite{Xu13}. 

In the waveguide, where both full and partial permutation symmetry are absent, we resort to approximate methods. The simplest is mean field, which neglects correlations by factorizing the density matrix and treating each emitter as a classical dipole. To include leading quantum corrections, we employ the Truncated Wigner Approximation (TWA) for dissipative spin systems, which yields $2N$ stochastic differential equations, where correlated white noise terms model quantum fluctuations~\cite{Polkovnikov10,Schachenmayer15,Mink22,Mink23,Guardiola25}. The reliability of this approach is supported by benchmarks against exact small-$N$ simulations, together with recent evidence that few-body observables in superradiant decay with all-to-all coupling are well captured by mean-field trajectories under heterodyne unraveling~\cite{Zhang25}. We use this method throughout the manuscript to assess the impact of quantum fluctuations.

\subsection*{Mean field and classical synchronization}
By assuming a separable state $\rho(t)\approx\bigotimes_{n=1}^N\rho_n(t)$, Eq.~\eqref{eq:master} reduces to the mean-field equations~\cite{Zhu15}
\begin{subequations}
\label{FullMeanFieldEqs}
\begin{align}
    \dot{s}^z_\ell&=-\sp_\ell\sum_{n\neq \ell}\sp_n\left(2 J_{n\ell}\sin(\phi_{n\ell })+\Gamma_{n\ell}\cos(\phi_{n\ell })\right)\nonumber\\&-(\Gamma'+\Gamma_{1D}+w)s^z_\ell+\frac{w-\Gamma'-\Gamma_{1D}}{2},\label{MeanFieldz}\\
     \dot{\sp_\ell}&=s^z_\ell\sum_{n\neq \ell}\sp_n\left(2J_{\ell n}\sin(\phi_{n \ell})+\Gamma_{\ell n}\cos(\phi_{n \ell})\right)\nonumber\\&-\frac{\Gamma'+w+\Gamma_{1D}}{2}\sp_\ell,\label{MeanFieldperp}\\
     \dot{\phi}_\ell&=\frac{s^z_\ell}{\sp_\ell}\sum_{n\neq \ell}\sp_n \left(-2 J_{\ell n}\cos(\phi_{n \ell})+\Gamma_{\ell n}\sin(\phi_{n \ell})\right)\label{MeanFieldphases}.
\end{align}
\end{subequations}
 Relative phases determine the coherent and collective dissipative evolution, as seen from the terms proportional to $J_{nm}$ and $\Gamma_{nm}$. In all mean-field simulations, we take the atoms to be fully excited at initial time and incorporate fluctuations by sampling initial conditions from the distribution $p(s^z,s^\perp,\phi)\propto e^{-2{s^\perp}^2}\delta\left(\sqrt{{\left(\sp\right)}^2+{\left(s^z\right)}^2}-\tfrac{1}{2}\right)$~\cite{Schachenmayer15}.
\bibliography{references}
\clearpage
\onecolumngrid
\setcounter{equation}{0}
\begin{center}
    {\large \bf{Supplementary Information
}}\\
\end{center}
\section{\label{sec:ansatzWG}Steady-state ansatz for the waveguide}

\begin{figure*} \centering \includegraphics[width=0.8\textwidth]{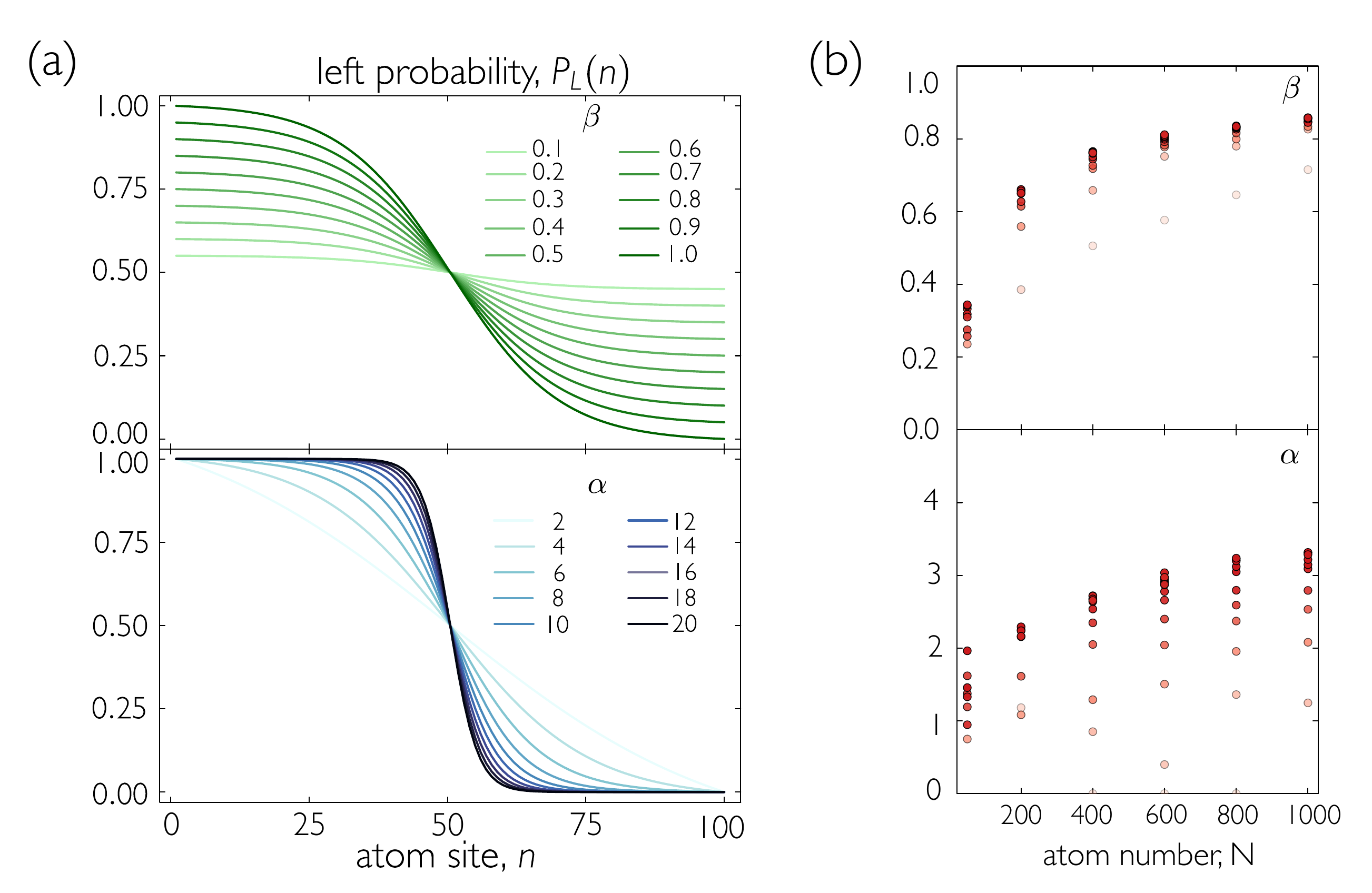}\caption{\textbf{Modeling the emergent phase patterns in the waveguide.} (a) Probability distribution $P_n(L)$ in Eq.~\eqref{eq:ProbModelSingleAtom} for the $n-$th atom locking to the left, shown for different values of $\beta$ (setting $\alpha=5$) and $\alpha$ (setting $\beta=1$). (b) Fitted values of the ansatz parameters $\alpha$ and $\beta$ for several system sizes and pumping rates, with fixed lattice constant $kd=2\pi/3$ and independent decay $\Gamma'=2\ga$. For every $N$, each point corresponds to a different pumping value $w$, and its opacity is proportional to the decay rate of that configuration relative to the maximal decay rate.}
\label{fig:Interpolations} \end{figure*}

 Here, we provide additional details on the steady-state ansatz in Eq.~\eqref{eq:ansatzWaveguide} of the main text, denoted as $\hat{\rho}_\text{ans,WG}(t_{ss})$. To fully specify the ansatz, one must provide the configuration probabilities $P(\mathbf{A})$. However, to compute the correlation matrix elements $C_{nm}$, only the marginals $P(A_n,A_m)$ are required. We propose a specific form for $P(A_n,A_m)$ that incorporates the fact that atoms on the left (right) side of the array preferentially synchronize to the left (right) order, while central atoms choose either with comparable probability. Concretely, we model the probability that atoms $n$ and $m$ synchronize to the same order as
\begin{subequations}\label{ansatz_probability_independent}
\begin{align}
    &P_{nm}(L,L)   = \gamma P_n(L)P_m(L),\\
    &P_{nm}(R,R)   = \gamma[1- P_n(L)][1-P_m(L)],
\end{align}
\end{subequations}
with 
\begin{equation}
    P_n(L) = \frac{1}{2} + \beta \frac{\tanh\left(\alpha \left(\frac{1}{2}-\frac{n-1}{N-1}\right)\right)}{2\tanh (\alpha/2)}, \label{eq:ProbModelSingleAtom}\\
\end{equation}

where $\beta\in[0,1]$ and $\alpha>0$ are parameters that control, respectively, how strongly atoms near the edges are pinned to their corresponding orders and how sharp the crossover is between the two domains near the array center. For values $\beta\approx 0$, the joint probability of two atoms is independent of their relative position, and the phase pattern of $\text{arg}[C_{nm}]$ obtained from $\hat{\rho}_\text{ans,WG}(t_{ss})$ matches the behavior characteristic of a ring cavity. In the opposite limit $\beta\approx1$, the atoms at the edges are fully locked to the corresponding order. Sample plots of this model for different parameters $(\alpha,\beta)$ are shown in Fig.~\ref{fig:Interpolations}(a).

A direct calculation shows that the ansatz leads to a correlation function of the form
\begin{equation}
\begin{aligned}
    &C_{nm}\propto P_{nm}(R,R)e^{-\ii k(z_n-z_m)}+ P_{nm}(L,L)e^{\ii k(z_n-z_m)}\\&+P_{nm}(R,L)e^{-\ii k(z_n+z_m)}\int_0^{2\pi}\frac{d\psi_R}{2\pi}\int_0^{2\pi}\frac{d\psi_L}{2\pi} e^{\ii(\psi_R-\psi_L)}\\&+P_{nm}(L,R)e^{\ii k(z_n+z_m)}\int_0^{2\pi}\frac{d\psi_R}{2\pi}\int_0^{2\pi}\frac{d\psi_L}{2\pi} e^{\ii(\psi_L-\psi_R)}.
    \end{aligned}
\end{equation}

The last two terms vanish due to the arbitrary relative phase between the two orders. Therefore, specifying the probabilities $P_{nm}(L,R)$ and $P_{nm}(R,L)$ is not required to determine the correlations. Their contribution is included indirectly through the parameter $\gamma\in[0,1]$, which reduces the probability that atoms $n$ and $m$ lock simultaneously to the same order. This suppresses $P_{nm}(L,L)$ and $P_{nm}(R,R)$ relative to the disordered configurations $[A_n = L, A_m = R]$ and $[A_n = R, A_m = L]$.

 The specific parameters of the model $(\alpha,\beta)$ are obtained by fitting the ansatz to the correlation phases extracted from the TWA numerical simulations. The fitted parameters for multiple system configurations are summarized in Fig.~\ref{fig:Interpolations}(b). They do not exhibit a strong dependence on the pumping strength. As $w$ approaches the optimal pumping rate $R_{max}$, the main effect is a reduction in the noise of the correlation phases $\arg[C_{(n+d)n}]$. Notably, the fitted values of $\beta$ tend to saturate as $N$ increases, approaching a value $\beta(N,\bar w^{(opt)}) \rightarrow \beta^* < 1$. This indicates that atoms near the edges do not become completely locked to their preferred order, and that some residual frustration persists even in very large arrays, with $N \sim 1000$.

We note that the ansatz in Eq.~\eqref{eq:ansatzWaveguide} is a phenomenological model intended to reproduce the behavior of correlations after averaging over trajectories. A more detailed analysis of individual trajectories may provide further insight on how quantum fluctuations influence the mean-field order. For example, although the relative phase between the left and right order parameters is assumed to be arbitrary, more general models could be developed by introducing a probability distribution of the relative phase, $P(\psi_R-\psi_L)$. The present ansatz reproduces the correct average correlations provided that the probability distribution of this relative phase is an even function. A possible refinement of the model would be to allow for atoms that are not synchronized with either order parameter. Nonetheless, the minimal model in Eq.~\eqref{eq:ansatzWaveguide} already demonstrates that the mean-field intuition is consistent with the TWA numerical results.

\section{\label{sec:Superspins}Large spin limit and thresholds}
Here we provide details of the analytic calculation of steady-state properties and thresholds for the ring cavity, Eqs.~\eqref{ThresholdsRing}.

The evolution of any spin operator \(\hat{\mathcal{O}}\) follows the Heisenberg equation 
\begin{equation}
    \frac{d\hat{\mathcal{O}}}{dt} = \ii\left[\hat{H},\hat{\mathcal{O}}\right]+\mathcal{L}^\dagger_\text{dis}[\hat{\mathcal{O}}]+\mathcal{L}^\dagger_\Gamma[\hat{\mathcal{O}}]+\mathcal{L}^\dagger_w[\hat{\mathcal{O}}] +\mathcal{F}\left[\hat{\mathcal{O}}\right],\label{Heisenberg}
\end{equation}
where the adjoints of the dissipators are
\begin{subequations}
\begin{align}
\mathcal{L}^\dagger_\text{dis}[\hat{\mathcal{O}}] &=\sum_{n,m=1}^N\frac{\Gamma_{nm}}{2}\left(2\heg^n\hat{\mathcal{O}}\hge^m-\hat{\mathcal{O}}\heg^n\hge^m-\heg^n\hge^m\hat{\mathcal{O}}\right), \\
\mathcal{L}^\dagger_\Gamma[\hat{\mathcal{O}}]&=\sum_{n=1}^N\frac{\Gamma'}{2}\left(2\heg^n\hat{\mathcal{O}}\hge^n-\hat{\mathcal{O}}\heg^n\hge^n-\heg^n\hge^n\hat{\mathcal{O}}\right),\\
\mathcal{L}^\dagger_w[\hat{\mathcal{O}}]&=\sum_{n=1}^N\frac{w}{2}\left(2\hge^n\hat{\mathcal{O}}\heg^n-\hat{\mathcal{O}}\hge^n\heg^n-\hge^n\heg^n\hat{\mathcal{O}}\right).
    \end{align}
\end{subequations}
Here $\hat{H}$ is the Hamiltonian defined in Eq.~\eqref{Ham}, and $\mathcal{F}\left[\hat{\mathcal{O}}\right]$ is quantum Langevin noise, which averages to zero and can be neglected when computing expectation values of $\hat{\mathcal{O}}$~\cite{Rubies23}.

As mentioned in Methods, for atoms coupled to a ring cavity, spacings satisfying $d= m\lambda_0 /2p$, with $m,p\in \mathbb Z$ exhibit partial permutational symmetry. We therefore represent each permutationally symmetric group $\mathcal{S}_\alpha$ ($\alpha=1,\dots,p$) by macroscopic angular-momentum–like operators~\cite{Lee25}
\begin{subequations}
\begin{align}
     \hat{J}^\alpha_-=&\frac{1}{N_\alpha}\sum_{n\in \mathcal{S}_\alpha} (-1)^{m(c_{n,\alpha}+1)}\hge^n,\\
    \hat{J}^\alpha_z=&\frac{1}{2N_\alpha}\sum_{n\in \mathcal{S}_\alpha} \hz^n,
\end{align}
\end{subequations}
where $c_{n,\alpha}=\lfloor (n-1)/\alpha \rfloor+1$ counts how many wavelengths atom $n$ is away from the first atom in its symmetry group, so that the factor $(-1)^{m(c_{n,\alpha}+1)}$ adds contributions within superspin $\mathcal{S}_\alpha$ in phase. Notice that because of the $1/N_\alpha$ normalization these operators do not form a $SU(2)$ algebra.

Evaluating Eq.~\eqref{Heisenberg} for these operators and applying a second-order cumulant expansion~\cite{Plankensteiner22,Rubies23}, the dynamics of $p$ incoherently driven superspins reads
\begin{subequations}\label{eq:ssAll}
\begin{align}
     \frac{d}{dt}\Jz{\alpha}&=-\sum_{\mu\neq\alpha}\frac{\Gamma_{\alpha\mu}N_\mu}{2}\left(\Jcd{\alpha}{\mu}\nonumber+\Jcd{\mu}{\alpha}\right)\\&-\ga N_\alpha \Jcs{\alpha}-(\gp+w)\Jz{\alpha}+\frac{w-\gp}{2},\label{eq:ss1}\\
     \frac{d}{dt}\Jcs{\alpha}&=\Jz{\alpha}\bigg[2\ga N_\alpha \Jcs{\alpha}+\sum_{\mu\neq \alpha}\Gamma_{\alpha\mu} N_\mu\big(\Jcd{\alpha}{\mu}\nonumber\\&+\Jcd{\mu}{\alpha}\big)\bigg]-(w+\gp)\Jcs{\alpha}+\frac{w}{N_\alpha},\label{eq:ss2}\\
     \frac{d}{dt}\Jcd{\alpha}{\xi}&=\ga\Jcd{\alpha}{\xi}\left(N_\alpha \Jz{\alpha}+N_\xi\Jz{\xi}\right)\nonumber\\&+\Gamma_{\xi\alpha}\left(N_\xi\Jcs{\xi}\Jz{\alpha}+N_\alpha\Jcs{\alpha}\Jz{\xi}\right)\nonumber\\&+\sum_{\mu\neq\alpha,\xi}N_\mu\left(\Gamma_{\mu\alpha}\Jcd{\mu}{\xi}\Jz{\alpha}+\Gamma_{\xi\mu}\Jcd{\alpha}{\mu}\Jz{\xi}\right)\nonumber\\&-(\gp+w)\Jcd{\alpha}{\xi}.\label{eq:ss3}
\end{align}
\end{subequations}

For $p=1$ and large $N$, Eqs.~\eqref{eq:ssAll} reduce to the superradiant laser equations of Ref.~\cite{Meiser10}, i.e.,
\begin{subequations}\label{eq:FullLaser}
\begin{align}
\frac{d}{dt} \langle\hat{J}_z\rangle &= -\ga N\langle\hat{J}_+\hat{J}_-\rangle -(\gp+w)\langle\hat{J}_z\rangle+\frac{w-\gp}{2},\\
\frac{d}{dt} \langle\hat{J}_+\hat{J}_-\rangle &=-(\gp+w)\langle\hat{J}_+\hat{J}_-\rangle +2N\ga \langle\hat{J}_z\rangle \langle\hat{J}_+\hat{J}_-\rangle.
\end{align}
\end{subequations}
The steady-state population and emission rate are in that case
\begin{subequations}
\begin{align}
       R(t=t_{ss})&=\begin{cases}
            0  \text{     if    } w\notin [\gp,N\ga] \\ 
            \begin{aligned}
                \frac{wN}{2}\left(1-\frac{\gp}{w}-\frac{1}{w}\frac{(w+\gp)^2}{N \ga}\right)\\[0.7em]
            \text{     if    } w\in [\gp,N\ga].
            \end{aligned}
        \end{cases}
        \label{sol1SL_laser}\\
         P_e(t=t_{ss})&=\begin{cases}
            \frac{N(w-\gp)}{2(w+\gp)} +\frac{N}{2}  \text{     if    } w\notin [\gp,N\ga] \\ 
            \frac{(w+\gp)}{2\ga}+\frac{N}{2} \text{     if    } w\in [\gp,N\ga].
        \end{cases}
        \label{sol2SL_laser}
\end{align}
\end{subequations}

For $p\neq1$, we assume all superspins contain the same number of atoms ($N_\alpha =N_s\text{  }\forall\text{  } \alpha$) and note that  $\Gamma_{\alpha\beta} = \ga\cos(kd(\alpha-\beta))$ is a circulant matrix, so the equations are invariant under any global cyclic permutation and:
\begin{enumerate}
    \item Since the steady state is unique, single-superspin variables are identical across $\alpha$ in the long time limit regardless of the initial state, i.e. $\Jz{\alpha}\equiv\Jz{1}$ and $\Jcs{\alpha}\equiv\Jcs{1}$ $\forall$ $\alpha$.
    \item Two-superspin correlations depend only on separation. In fact, for all $t$ and $\alpha\neq\beta$, $\Jcd{\alpha}{\beta}(t)=\Gamma_{\alpha\beta}C(t)$, i.e., correlations between different superspins are described by a global function weighted by the coupling. 
\end{enumerate}

Harnessing these observations and the identity
\begin{equation}
    \sum_{\mu=1}^p \Gamma_{\alpha\mu}\Gamma_{\mu\xi} = \frac{p}{2}\ga\Gamma_{\alpha\xi},
\end{equation}
the system \eqref{eq:ssAll} reduces from $p+p^2$ equations to two by defining $R_\alpha = \sum_\mu \frac{\Gamma_{\mu\alpha}}{\ga}\Jcd{\mu}{\alpha}$:
\begin{subequations}\label{eq:FullRingLaser}
\begin{align}
\frac{d}{dt} \Jz{\alpha} &= -\ga N_sR_\alpha -(\gp+w)\Jz{\alpha}+\frac{w-\gp}{2},\\
\frac{d}{dt} R_\alpha &=-(\gp+w)R_\alpha +N\ga \Jz{\alpha}R_\alpha.
\end{align}
\end{subequations}

Equations~\eqref{eq:FullRingLaser} are thus the ring-cavity analogue of \eqref{eq:FullLaser}, differing only by the missing factor of two in the nonlinear term for $R_\alpha$ and by the replacement $N\to N_s$ in the inversion equation. The resulting steady state corresponds to Eqs.~\eqref{ThresholdsRing} in the main text.

\section{\label{sec:AppendixLinewidth_formerMainText}Spectral properties of emitted light}

Using an extension of the TWA to compute multi-time correlators~\cite{Guardiola25} together with an $SU(4)$ representation of the Liouvillian for the single-mode cavity~\cite{Xu15}, we now discuss the spectral properties of light emitted from a ring cavity and a waveguide, and numerically explore the possibility of linewidth narrowing with increasing atom number.

For atoms coupled to a single-mode cavity, the two-time correlation of any pair of atoms decays exponentially with the time delay $\tau$, i.e.,
 \begin{equation}
\langle\hge^n(t_{ss}+\tau)\heg^m(t_{ss})\rangle\sim e^{-\tau/t_\text{coh}},
\end{equation}
with a coherence time $t_\text{coh}$ that increases with $N$~\cite{Meiser10}. For large atom number, the spectrum of the emitted cavity field is
\begin{equation}
\scalebox{0.90}{$
S_{\hat{E}^+_C}(\omega) \simeq N(N-1)\,\text{Re}\!\left[\int_0^\infty d\tau\, e^{-\ii\omega \tau}
\langle \hge^1(t_{ss}+\tau)\heg^2(t_{ss})\rangle\right]$}.
\label{EqSpectrum}
\end{equation}
The long-lived correlations cause $S_{\hat{E}^+_C}(\omega)$ to develop a linewidth approaching $(\Delta\nu)_\text{min}\sim\ga$, which no longer reflects the phase noise introduced by individual incoherent processes such as parasitic decay or incoherent pumping. The residual broadening is primarily due to phase fluctuations of the emitted field~\cite{Xu16}: the amplitude is fixed while the phase remains undetermined. Reduced phase noise enables such light to be used to lock a laser, or as an active frequency reference.

\begin{figure}[t]
\centering
\includegraphics[width=0.5\columnwidth]{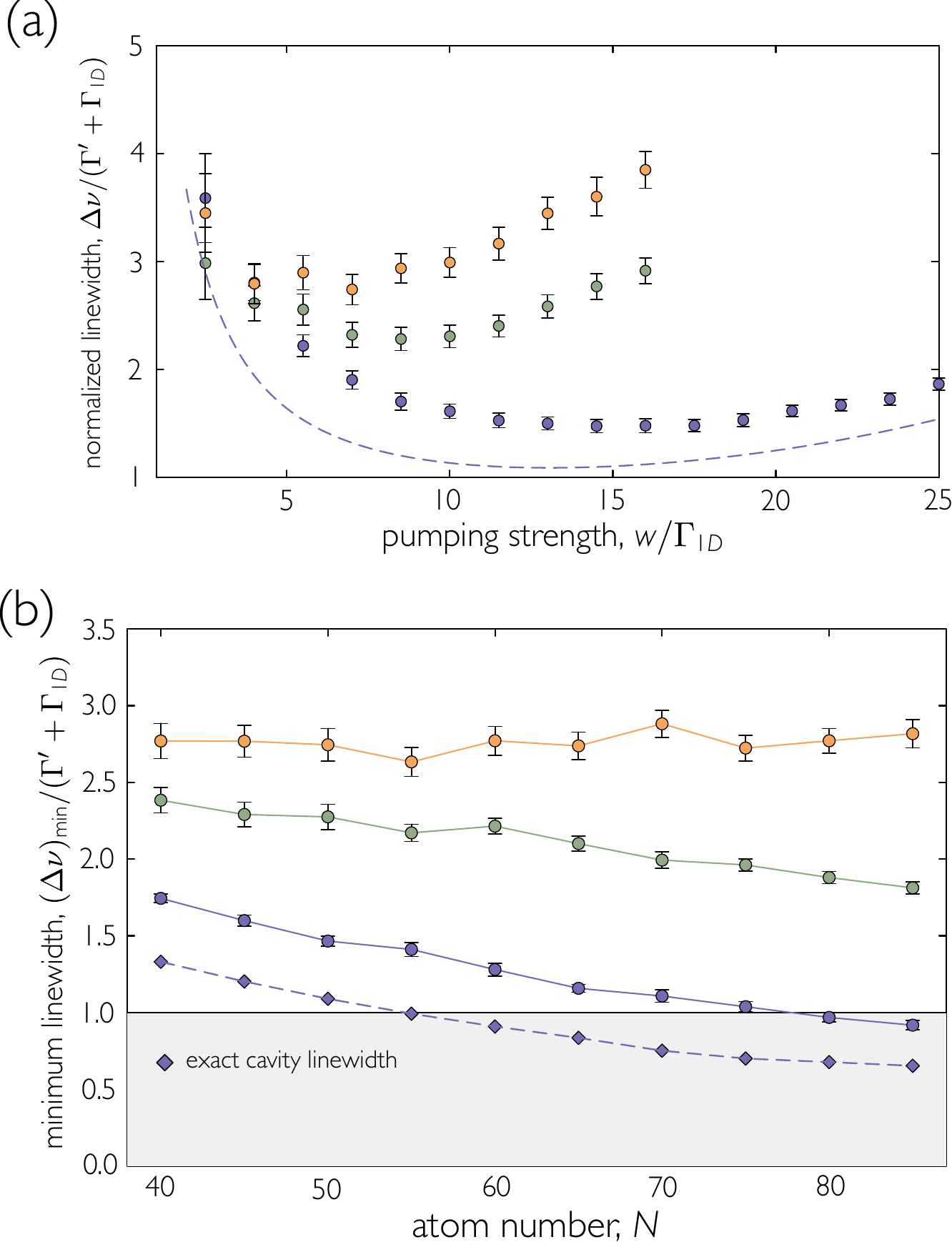}
\caption{\textbf{Linewidth of the emitted field.} (a) Linewidth vs. pumping strength for $N=50$ atoms for the single-mode cavity, ring cavity and waveguide models. (b) Scaling of the linewidth vs $N$. In both panels, points are obtained using the TWA, dashed lines show exact results based on the $SU(4)$ formalism, and solid lines serve as guides to the eye. For the ring cavity and waveguide, $(kd,\gp)=(2\pi/3,2\ga)$.}
\label{fig:LinewidthFarField}
\end{figure}

Figure~\ref{fig:LinewidthFarField} summarizes the linewidth behavior as a function of pump strength and atom number for the three spin models discussed in the main text. For moderate $N$, the linewidth exhibits a non-monotonic dependence on the pump rate within the synchronization window [Fig.~\ref{fig:LinewidthFarField}(a)], revealing an optimal pump rate that minimizes the linewidth. Linewidths are computed using TWA~\cite{Guardiola25} and validated against exact $SU(4)$ calculations for the cavity, which reproduce the same qualitative trends.

Figure~\ref{fig:LinewidthFarField}(b) shows the minimum linewidth as a function of atom number. Both the spectral linewidth of $\hat{E}^+_C$ in the single-mode cavity and that of $\hat{E}^+_{RC,R/L}$ in the ring cavity exhibit a seemingly downward trend with increasing $N$. In the waveguide, we calculate the linewidth of the right (left) field at an observation point to the right (left) of all atoms. After accounting for the estimation uncertainty (see below for more details on the analysis), our numerical results do not conclusively demonstrate linewidth narrowing. This inconclusiveness may arise from numerical limitations. For instance, comparison with exact linewidths in the cavity case, obtained using an $SU(4)$ representation of the dynamics~\cite{Xu13}, shows that the TWA captures the true linewidth narrowing only qualitatively. In addition, narrowing may only emerge at atom numbers beyond our current simulation capabilities. Alternatively, the absence of clear narrowing may be a genuine physical effect, indicating that the spin order in the waveguide is not robust enough to eliminate noise from the parasitic decay.

The seemingly downward trend of the linewidth for the ring cavity is consistent with the picture developed above, where the atoms synchronize to the order set by either $\hat{\mathcal{O}}_{R}$ or $\hat{\mathcal{O}}_{L}$. Although these operators no longer correspond to a single macroscopic dipole, they still represent a global phase order that becomes increasingly robust as more atoms participate, since collective feedback suppresses phase fluctuations of individual atoms. The resulting long-lived phase memory is thus directly visible in the far-field spectra. Nonetheless, we remark that true line narrowing (i.e., $\Delta\nu<\Gamma'+\ga$) is not achieved within the system sizes accessible to our numerics, although both the observed trends and the nature of the synchronized order strongly suggest that such narrowing should occur in the large-$N$ limit. This indicates that superradiant lasing should be observable in a ring cavity, despite mode competition.

\subsection{\label{sec:AppendixLinewidth}Linewidth analysis}

Here, we provide further details on the calculation of the linewidth, the scaling of the minimum linewidth with the atom number $N$, and the estimation of error bars. We also discuss the main limitations of the numerical techniques employed.

In the absence of symmetries, the numerical complexity for computing the linewidth of any field using the TWA scales as $\mathcal{O}(M N^3)$~\cite{Guardiola25}, where $M$ is the number of trajectories required for convergence. To estimate the linewidths shown in Fig.~\ref{fig:LinewidthFarField}(a) of the main text, we compute the two-time correlator of either the cavity field or the right-propagating field and fit it to an exponential decay of the form
\begin{equation}
    \langle \hat{E}^-(t_{ss}+\tau)\hat{E}^+(t_{ss})\rangle \sim e^{-\tau\Delta\nu/2}
\end{equation}

Although fluctuations inherent to the TWA induce a noise floor at late times, preventing the correlator from vanishing completely as $\tau\rightarrow \infty$, we find that the fitting procedure reliably captures the initial exponential decay of the correlator. The error bars in Fig.~\ref{fig:LinewidthFarField}(a) arise from the stochastic nature of the TWA simulations and correspond to three standard deviations estimated from the fitting procedure.\\

To determine the minimum linewidth, these calculations are repeated for several values of the incoherent pumping rate $w$. To reduce computational cost and mitigate the impact of noisy or outlier realizations, we do not perform a dense scan in $w$. Instead, we evaluate the linewidth at 10 to 20 selected pumping strengths, which are not necessarily uniformly spaced and may not include the value of $w$ that yields the absolute minimum linewidth. To estimate the minimum from these discrete samples, we fit the data to a LogSumExp function, $\frac{1}{p_0}\log\left(\exp{(p_0(p_1 w+p_2))}+\exp{(p_0(p_3 w+p_4))}\right)$, from which we obtain the minimum linewidth. Alternative fitting models, including polynomials and other non-linear functions, can be used, resulting in similar estimates for the minima. The error bars in Fig.~\ref{fig:LinewidthFarField}(b) likewise represent three standard deviations.\\

The application of these techniques to our data suggests that the approach is generally valid; however, the results reveal inherent noise that hinders robust conclusions. Because of the various fitting procedures and underlying assumptions involved, the method could be refined at different levels of the calculations, making further numerical studies necessary to assess the validity of this analysis and to extend our results to larger system sizes.

\end{document}